\documentclass[12pt]{iopart}
\usepackage{iopams, amssymb,amstext, graphicx}
\newcommand{\field}[1]{\mathbb{#1}}

\newcommand{\R}{\field{R}}

\begin{document}
\title{Effective Viscosity of Dilute Bacterial Suspensions: A Two-Dimensional Model}
\author{Brian M.~Haines$^1$, Igor S.~Aranson$^2$, Leonid Berlyand$^1$, Dmitry A.~Karpeev$^3$}

\address{$^1$ Department of Mathematics, Pennsylvania State
University, 418 McAllister Building, University Park, PA 16802 }
\address{$^2$ Materials Science Division, Argonne National
Laboratory, 9700 South Cass Avenue, Argonne, IL 60439}
\address{$^3$ Mathematics and Computer Science Division, Argonne
National Laboratory, 9700 South Cass Avenue, Argonne, IL 60439}

\begin{abstract}
Suspensions of self-propelled particles are studied in the framework of two-dimensional (2D)
Stokesean hydrodynamics.  A formula is obtained for the effective viscosity of such suspensions
in the limit of small concentrations.  This formula includes the two terms that are found in the 2D
version of Einstein's classical result for passive suspensions.  To this, the main result of the paper
is added, an additional term due to self-propulsion which depends on the physical
and geometric properties of the active suspension.  This term explains the experimental
observation of a decrease in effective viscosity in active suspensions.
\end{abstract}
\pacs{87.16.-b, 05.65.+b, 87.17.Jj}
%\submitto{\PB}
\section{Introduction}

Recently there has been considerable interest in understanding the dynamics of
systems of active interacting biological agents (active systems, for short),
such as flocking birds, schooling fish, swarming bacteria, etc \cite{wu,kim,mendelson,dombrowski,riedel,becco}.
Properties of these
strongly self-organizing  dissipative systems are of fundamental interest to nonequilibrium statistical
dynamics \cite{feder,toner,gregoire,vicsek,simha} and to potential technological applications \cite{kim1}.
%From the latter point of view
%natural active systems exhibit behavior prototypical of the future synthetic biomaterials with
%striking new properties.  In order to be able engineer such new materials with prescribed
%characteristics there is a need to quantify the effective emergent properties of active systems
%and their dependence on the details of the interparicle interactions.

Common swimming (motile) bacteria, such as {\it Bacillus Subtilis}, {\it Escherichia coli}, and many others, are rod-shape
microorganisms (length  about 5 $\mu m$, diameter of the order 1 $\mu m$), propelled by the input of mechanical energy
at the smallest scales by the rotation of helical flagella attached to the cell wall.
Suspensions of swimming bacteria (active suspensions) are a convenient representative of self-organizing biological systems. At relatively high filling fractions of  bacteria, they interact mostly through hydrodynamic entrainment induced by their swimming with respect to ambient fluid \cite{dombrowski,sokolov+}, which are instrumental in the establishment of
long-range order.  At the same time, they are amenable to accurate experimental studies, control,
and manipulation \cite{sokolov+}.
In the dense regime, the dynamics of active suspensions is dominated by the
multiple body interactions and long-range self-organized coherent structures, such as recurring whorls and jets with the spatial scale exceeding the size of individual bacterium by an order of magnitude.  This, however, makes the analysis rather challenging.

In the dilute regime, the hydrodynamic interactions between bacteria (or passive particles) are often ignored as the interparticle
distance exceeds the range of the flows resulting from the particle motion.
Therefore, in this case it is possible to isolate the effect of the particle-fluid interactions
on the effective properties of active suspensions.  This constitutes a step towards
the full understanding and utilization of the novel properties of active suspensions.

The first step was taken in \cite{einstein} by Einstein,
who derived an explicit formula for the effective
viscosity of a dilute suspension of passive spheres.  This formula shows an increase of the viscosity over that of the ambient fluid
alone.  The correction to the effective viscosity is to the first order in
the volume fraction of the inclusions.  Later, Batchelor and Greene obtained the second order asymptotic formula, which takes into account
pairwise interactions between particles \cite{batchelor1}.
Jeffrey calculated the effective viscosity of a suspension
of ellipsoidal particles to first order in \cite{jeffery}.
Unlike the spherical case, here the value of the viscosity
is affected by the distribution of orientations of the inclusions, which is assumed to be uniform.

In this work, Einstein's classical dilute limit result is extended to the case of
\emph{self-propelled} disk inclusions in a 2D Stokesean fluid.
The choice of two-dimensional hydrodynamics is motivated by the tractable nature of calculations involving
Green's functions in 2D and also by the quasi-two-dimensional thin film geometry of the experiment \cite{sokolov+}.
In particular, the correction to the effective
viscosity is explicitly calculated as a function of the orientations of the bacteria's flagella, the intensity
of the their force of self-propulsion, and the volume fraction of the suspension.
In the case of an absence of self propulsion and, alternatively, the case of a uniform distribution of orientations,
the result recovers the 2D version of Einstein's result (see, e.g.\cite{belzons}).

The model of the bacterium and the corresponding hydrodynamics are described in
Section~\ref{sec-model}.  Following Batchelor \cite{batchelor}, the effective viscosity of a dilute
suspension is defined in Section~\ref{sec-model}, in terms of a suitable background flow and the
disturbance flow produced by the inclusion of a bacterium in the background flow.  In the case of
self-propulsion the disturbance flow is due both to the passive response of the background flow to
the inclusion of a particle as well as the response of the stationary fluid to the particle's self-propulsion.
Because of the linearity of Stokesean hydrodynamics, the two components of the disturbance
flow can be computed independently using the Green's function for the disk.

The disturbance flow $u_p^\prime$ due to the locomotion of a single bacterium is calculated in Section~\ref{sec-flow}.  This
flow translates the bacterium.  It is also shown that this disturbance flow decays at infinity.
This result is emphasized since it is a point of well-known deficiency of 2D hydrodynamics.
Indeed, while the translation of a ball in 3D produces no flow at infinity, in the well-known Stokes paradox,
the 3D flow due to an infinite cylinder
(described by the corresponding force monopole) translating transversely to its axis generates
a non-zero flow at infinity (see, e.g., \cite{landau-lifshitz}).  Thus  prevents  the decoupling of rods,
which is needed for the dilute limit, where it is assumed that the flow due to a suspension can be approximated
by the sum of solutions due to a single inclusion.
However, a self-propelled particle is constrained by the viscous drug force opposing the propulsion force, thus producing a force dipole
which decays at infinity, unlike the force monopole in the case of the moving disk.

For the other disturbance flow, $u_{int}'$, due to the bacterium's passive response to a background flow,
care must be taken to ensure that no flow is produced at infinity.  Thus, in Section~\ref{sec-ev}
such a flow is selected to ensure the disturbance decays at infinity.  Once it has been done,
the effective viscosity is calculated as a function of the flagella's orientations relative to the
background flow. The time-dependent nature of the effective viscosity, due to active alignment of
the inclusions to the flow, is also discussed.  Conclusions are discussed in Section~\ref{sec-conclusion}.  Finally,
the details of calculations are discussed in the appendices.

\section{Model and its Homogenization}\label{sec-model}
\begin{figure}[h]
	\centering
		\includegraphics{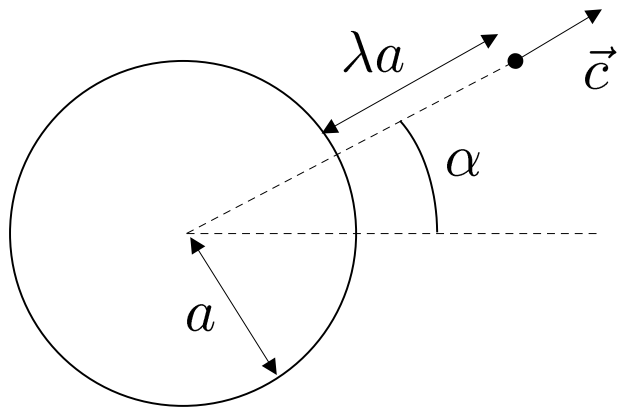}
	\caption{Schematic presentation of a bacterium with orientation $\alpha$.}
	\label{orientation}
\end{figure}

A 2D dilute suspension of bacteria is modeled as a collection of discs of radius $a$, each of which has an associated
point force, representing the flagellum,  placed at a distance $\lambda a$ from the disk, as shown in Figure~\ref{orientation}.  The point force is
directed radially outward from the center of the bacterium and has some orientation angle $\alpha$, measured from the
$x$-axis.  The bacteria are distributed throughout an ambient fluid of viscosity $\eta$ which takes up the entire domain
$\R^2$. A single bacterium in an infinite fluid moves with respect to the fluid with constant velocity driven by the
 point force $\vec c$.

The suspension is assumed to contain sufficiently many bacteria so as to produce appreciable changes in the properties of
the equivalent homogenized fluid.
At the same time, the size of the bacteria and the propulsion force are assumed to be sufficiently small, so that the disturbance
flow produced by bacteria  is negligible at inter-bacterial distances and hence can be ignored at the locations of other bacteria.
Therefore, in the following calculations, it is sufficient to consider a single bacterium with orientation $\alpha$
in an unbounded volume of fluid.  This is analogous to Einstein's assumptions for passive suspensions.

This unbounded volume is an idealization of the microscopic volume element
surrounding a single one of the particles located within a macroscopic fluid element.  Because of the
decay assumption, this idealization is permissible for the solution of the Stokes' equation within that region. The viscous
dissipation in the macroscopic fluid element is then the sum of the dissipations in all of the subordinate microscopic
volumes, containing a collection of particles having some prescribed distribution of orientations.
Analogous to the method of Batchelor in \cite{batchelor}, the effective viscosity $\eta^*$ of a suspension in an ambient fluid of
viscosity $\eta$ is defined as the viscosity
of an equivalent fluid with no inclusions that produces the same energy dissipation in each macroscopic volume element.

For a suspension of passive particles, the existence of an equivalent Newtonian fluid with the scalar effective viscosity
$\eta^*$ is a classical result of homogenization theory (see e.g., \cite{levpal} and references therein).  The analysis for
active particles is more subtle and has not been carried out in a rigorous mathematical context.  The assumption is made that the effective viscosity
can be defined in the same way for active particles.
Nevertheless, the technique is presented somewhat cryptically in the literature.  To clarify the main points,
the conceptual side of the calculation is presented here in some detail;
the technical details can be found in \ref{app-def_ev} 
(also see \cite{batchelor} for the derivation in the case of passive particles).

Since the effective viscosity is defined in terms of energy dissipation, a background flow (which describes the flow throughout
the homogenized fluid and on the boundary of the suspension) is chosen so that it has a non-zero rate of energy dissipation
(i.e., requiring a non-zero strain rate).
Choosing $\R^2$ to represent the microscopic volume with coordinates
$\{x_i\},\ i=1,2$ the velocity components of the background flow are $u_i=\epsilon_{ij}x_j$,
where $\epsilon_{ij}$ is constant and symmetric strain rate tensor  but otherwise left to be specified later.
Adding a single self-propelled disk inclusion $V_b$ somewhere
in the plane produces the disturbance flow $\vec{u}^\prime$.  Provided that $\vec{u}$ is selected properly, $\vec{u}^\prime$ vanishes at
infinity, decoupling the bacteria, as required of the dilute limit.  The disturbance flow can be computed explicitly and
it is done below.  The total flow of the suspension is $\vec{\tilde u} = \vec{u} + \vec{u}^\prime$ and in principle it should be possible
now to compare the dissipation rates of $\vec{u}$ and $\vec{\tilde u}$ by integrating over $\R^2$ and $\R^2 - V_b$
respectively.

\begin{figure}[h]
	\label{fig-Omega}
	\centering
		\includegraphics[scale=0.50]{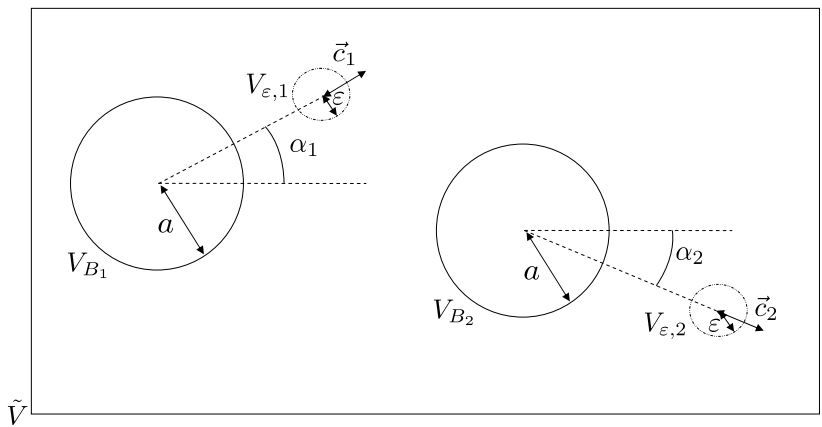}
	\caption{Bounded domain $\tilde V$ for which the modified dissipation rate is computed; includes the complement of two
$a$-disks (bacteria) and $\varepsilon$-disks (neighborhoods of point forces).}
\end{figure}

As shown in \cite{batchelor}, the integration over the unbounded volume $\R^2$ leads to technical difficulties.  This motivates
the more efficient method of comparing the dissipation rates within a bounded domain $\tilde{V}$ which contains sufficiently many bacteria.
Figure~\ref{fig-Omega} shows such a domain encompassing two bacteria.
Inside $\tilde{V}$ are the regions representing the bodies of the bacteria, $V_{B_l}$ and, for each, an $\varepsilon$-disk $V_{\varepsilon,l}$ around the
corresponding point force. The disturbance flow in the domain $\tilde V - \sum_l V_{B_l} - \sum V_{\varepsilon,l}$ is required to vanish
at the outer boundary $\tilde \Gamma = \partial \tilde V$.
This models a finite container with the background flow applied at the boundary
and simplifies the computation of the dissipation rate integrals by replacing a volume integral with an integral
over the boundaries of the bacteria.

Calculating the disturbance flows $\vec{u}'$, however, involves solving a boundary value problem in
the two-dimensional domain $\tilde V - \sum_l V_{B_l} - \sum V_{\varepsilon,l}$.
To avoid this, the total dissipation rate in this domain can be equivalently calculated as the
integral of the work of forces at the boundary $\tilde \Gamma$.
 By requiring that $u^\prime$ vanish at $\Gamma=\partial V$, the flow of the
homogenized fluid  and total flow of the suspension agree on the boundary $\Gamma$, $u_i = \tilde u_i = \epsilon_{ij}\,x_j$,
but the corresponding values of the stress tensors $\sigma$ and $\tilde \sigma$ are not necessarily the same.  Now, the density of work at the boundary
is $n_i \sigma_{ij}\epsilon_{jk} x_k$ and $n_i \tilde \sigma_{ij}\epsilon_{jk} x_k$
for the homogenized and total suspension flows respectively.  The effective viscosity is defined by integrating these quantities
over $\Gamma$ and setting them equal:
\begin{equation}
\int_\Gamma \sigma_{ij} u_i n_j dA = \int_\Gamma \tilde{\sigma}_{ij} \tilde{u}_i n_j dA.
\end{equation}
Using the Stokes equation $\nabla \cdot \sigma^\prime = 0$ and repeatedly applying the divergence theorem (see appendix \ref{app-def_ev}), the integrals of
these densities over $\tilde \Gamma$ can be reduced to integrals over the boundary of the bacterial domain only:
$\Gamma_b + \Gamma_\varepsilon = \partial V_b + \partial V_\varepsilon$.  These integrals are independent of $\tilde \Gamma$
and upon passing to the limit $\tilde V \longrightarrow \R^2$ they are expressed in terms of the disturbance
flow $\vec{u}^\prime$ vanishing at infinity.

The effective viscosity is then determined from the expression resulting from summing over all inclusions:
\begin{eqnarray}\label{origevintegral}
\lefteqn{2(\eta^*-\eta)\epsilon_{ij}\epsilon_{ij}} \\
&=\frac{\epsilon_{ik}}{A}\sum_l\int_{\partial V_{B_l}\cup \partial V_{\varepsilon,l}}\left(\sigma^\prime_{ij}x_k n_j-2\eta u^\prime_i n_k\right)dA. \nonumber
\end{eqnarray}
Applying the dilute assumption, the solution to (\ref{stokesfull}) is replaced by a superposition of solutions for the flow due to a single bacterium, derived in the next section.  Thus, for aligned bacteria, the above integral can be replaced with one over the boundary of one bacterium in $\R^2$,
\begin{eqnarray}\label{evintegral}
\lefteqn{2(\eta^*-\eta)\epsilon_{ij}\epsilon_{ij}=} \\
&\epsilon_{ik}\phi\int_{\partial V_{B}\cup \partial V_{\varepsilon}}\left(\sigma^\prime_{ij}x_k n_j-2\eta u^\prime_i n_k\right)dA, \nonumber
\end{eqnarray}
where $\phi$ is the volume fraction occupied by bacteria.

\section{Flow due to a single bacterium}\label{sec-flow}
Let $\Omega$ be the exterior of the unit disk in $\R^2$ and $\delta_{\vec{r}_0}$ be the Dirac delta evaluated at $\vec{r}-\vec{r}_0$, where $\vec{r}_0$ is the location of the point force.  In the low Reynolds number limit, the flow due to the bacterium obeys the Stokes equation
\begin{equation}\label{stokes}
\left\{
\begin{array}{rl}
\eta\Delta\vec{u_p'}&=\nabla p_p'+\vec{c}\delta_{\vec{r}_0} \text{ in } \Omega \\
\nabla\cdot\vec{u}_p'&=0 \text{ in } \Omega \\
\vec{u}_p'&=0 \text{ on } \Gamma
\end{array},
\right.
\end{equation}
where $\vec{c}$ is the force strength and $\Gamma=\partial\Omega$.  For simplicity of calculations, set $\vec{c}=c \hat{e}_1$ and $\vec{r}_0=(1+\lambda)\hat{e}_1$.
Note that here, $\vec{u}_p'$ is set to $0$ on $\Gamma$.  This is the reference frame in which the bacterium is moving--for the purpose of finding the total flow
of the suspension, the reference frame in which the fluid is at rest at infinity must be used.
%Note that no-slip boundary conditions are not
%enforced at the location of the point force ($\vec{r}_0$).  Thus, by the linearity of the Stokes equation, any other
%hydrodynamic interactions will be the same as with a disk.

By writing $\vec{u}$ as the curl of a stream function $\Phi$ and taking the divergence of (\ref{stokes}), this reduces to the inhomogeneous biharmonic equation
\begin{equation}\label{biharmonic}
\left\{
\begin{array}{c}
\Delta^2\Phi(x,y)=\frac{\partial}{\partial y} \frac{c}{\eta}\delta_{\vec{r}_0} \text{ in } \Omega \\
\Phi=\frac{\partial\Phi}{\partial n}=0 \text{ on } \Gamma
\end{array}.\right.
\end{equation}

The Green's function $G$ for the domain $\Omega$ with these boundary conditions is known and yields the solution
\begin{equation}
\Phi(x,y)=-\frac{c}{\eta}\frac{\partial G}{\partial \gamma}(x+iy,1+\lambda).
\end{equation}
Explicitly, in polar coordinates,
\begin{eqnarray}
%\begin{split}
\lefteqn{\Phi(r,\theta)=} \\
&\frac{cr\cos(\theta)}{8\pi\eta}\left[\frac{((1+\lambda)^2-1)(r^2-1)}{1+(1+\lambda)^2r^2-2(1+\lambda)r\cos(\theta)}\right. \nonumber \\
&\left.+\log\left(\frac{(1+\lambda)^2+r^2-2(1+\lambda)r\cos(\theta)}{1+(1+\lambda)^2r^2-2(1+\lambda)r\cos(\theta)}\right)\right]. \nonumber
%\end{split}
\end{eqnarray}
\begin{figure}
	\centering
		\includegraphics[width=.4\textwidth]{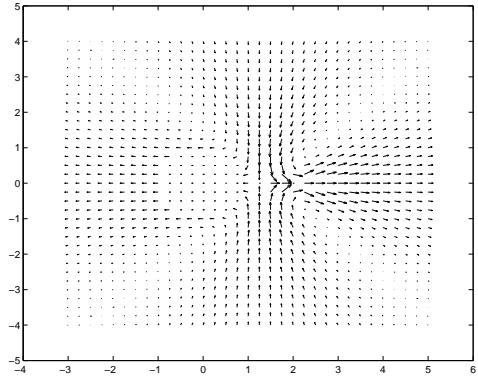}
	\caption{Velocity field produced by a bacterium in 2D geometry.}
	\label{velocity}
\end{figure}

It is easy to see that $u_p'\rightarrow u_\infty=\text{const.}$ as $r\rightarrow\infty$.  Subtracting this flow
(moving to the reference frame where the water is at rest at $\infty$) and rescaling to a disk of radius $a$, is shown in figure \ref{velocity} and is given asymptotically by
\begin{eqnarray}
\lefteqn{\vec{u}(r,\theta)=} \\
&\left(
\begin{array}{c}
\frac{ca^2\lambda^2(2+\lambda)^2}{8\pi\eta(1+\lambda)^3}\frac{1}{r}(\cos(\theta)+\cos(3\theta))+O(\frac{1}{r^2}) \\
\frac{ca^2\lambda^2(2+\lambda)^2}{8\pi\eta(1+\lambda)^3}\frac{1}{r}(\cos(2\theta)\sin(\theta))+O(\frac{1}{r^2})
\end{array}\right). \nonumber
\end{eqnarray}

Now, by taking the divergence of (\ref{stokes}), one obtains
\begin{equation}\label{pressurepde}
\left\{
\begin{array}{rl}
\Delta p&=-c\frac{\partial}{\partial x}\delta_{\vec{r}_0} \text{ in } \Omega\\
\nabla p\cdot\hat{n}&=\eta\Delta\vec{u}\cdot\hat{n} \text{ on } \Gamma
\end{array}.
\right.
\end{equation}
The solution to (\ref{pressurepde}) is found using complex function theory (see appendix \ref{app-pressure}), and is the real part of the complex function
\begin{equation}
\label{analyticpressure}
\Pi(z)=-\frac{c}{2\pi}\frac{a^2\lambda^2(2+\lambda)^2z}{(1+\lambda)(a(1+\lambda)-z)((1+\lambda)z-a)^2}.
\end{equation}

\section{Calculating the effective viscosity}\label{sec-ev}
In order to calculate the effective viscosity, it remains to perform the integration in (\ref{evintegral}).  The effective viscosity is independent of the background flow, so one is free to choose the flow which simplifies the necessary calculations.  In (\ref{evintegral}), $\vec{u}^\prime$ and $p^\prime$ are the solutions to
\begin{equation}
\left\{
\begin{array}{l}
\eta\Delta\vec{u}'=\nabla p'+\vec{c}\delta_{\vec{r}_0} \\
\nabla\cdot\vec{u}'=0
\end{array}\right.
\end{equation}
subject to the boundary conditions
\begin{equation}\label{bgflow-bcs}
\begin{array}{ll}
u^\prime_i\rightarrow \epsilon_{ij} x_j & \text{as } r\rightarrow\infty \\
\vec{u}^\prime=const & x\in\Gamma.
\end{array}
\end{equation}
Note that, as mentioned in section \ref{sec-flow}, one is not free to choose the value of the constant that $\vec{u}$ takes on $\Gamma$.

The formula for the effective viscosity does not depend on the choice of background flow (i.e., the choice of the strain rate tensor $\epsilon_{ij}$ in (\ref{bgflow-bcs})).  However, the choice of this flow is not completely arbitrary.  First, since the effective viscosity is defined in terms of energy dissipation, a non-zero rate of strain is required so that there is energy being dissipated due to viscosity.  Second, the background flow is chosen so that it does not rotate bacteria, which allows for the computation of the effective viscosity for a fixed (time-independent) distribution of orientations of bacteria (see figure \ref{orientation}).  Third, upon placing a disk (i.e., bacterium) in the flow, it is necessary that the disturbance flow $\vec{u}_{int}^\prime \rightarrow0$ as $r\rightarrow\infty$.  If this were not the case, the dilute assumption that this flow is negligible at the locations of other bacteria would be violated.  In 2D, this is only possible for a background flow that satisfies the conditions for existence of such a decaying flow derived in \cite{circlethm}.  The background flow is chosen to be the simplest flow that matches these criteria--the sum of two perpendicular shearing flows,
\begin{equation}
\label{bgflow}
\epsilon=2\epsilon_0\left(
\begin{array}{cc}
0&1\\ 1&0 \end{array}
\right).
\end{equation}
where $\epsilon_0=const $ is the amplitude of the strain rate.

The disturbance flow due to the presence of a bacterium will be
\begin{equation}
\vec{u}^\prime =\vec{u}_p^\prime (\alpha)+\vec{u}_{int}^\prime
\end{equation}
with the pressure given by
\begin{equation}
p^\prime =p_p^\prime(\alpha)+p_{int}^\prime
\end{equation}
where, using notation, $\vec{u}_p^\prime(\alpha):=\vec{u}_p^\prime(r,\theta-\alpha)$ as a polar vector and $p_p^\prime(\alpha):=p_p^\prime(r,\theta-\alpha)$, rotated to allow for an arbitrary orientation of a bacterium in the background flow.

For a suspension of bacteria that are not fully aligned, the effective viscosity can be written in terms of the function $X(\alpha)$ which gives the probability density of orientations of the bacteria $\alpha_l$, which are assumed to be independent identically distributed random variables.  In this case, (\ref{evintegral}) becomes
\begin{eqnarray}
\lefteqn{2(\eta^*-\eta)\epsilon_{ij}\epsilon_{ij}=} \\
&\epsilon_{ik}\phi\int_0^{2\pi}\int_{\partial V_{B}\cup \partial V_{\varepsilon}}\left(\sigma^\prime_{ij}(\alpha)x_k n_j \right.\nonumber \\
&\left. -2\eta u'_i(\alpha) n_k\right)dA X(\alpha)d\alpha, \nonumber
\end{eqnarray}
where $\phi$ is the volume fraction occupied by bacteria.
Performing the integration over $\partial V_B\cup \partial V_\varepsilon$, one obtains the following formula for the effective viscosity:
\begin{eqnarray}\label{evformula}
\lefteqn{\eta^*=\eta\left[1+2\phi\right.} \\
& \left.+\frac{c}{2\pi\eta\epsilon_0}\frac{1+2\lambda(2+\lambda)}{(1+\lambda)^3}\phi\int_0^{2\pi}\sin \left(2\alpha\right) X(\alpha)d\alpha\right], \nonumber
\end{eqnarray}
where $\phi$ is the two-dimensional volume fraction.

Note that for isotropically  oriented bacteria ($X(\alpha)=1/2 \pi=const$), the angular term goes to zero and the effective viscosity reduces to the result in \cite{belzons} for the effective viscosity of a two-dimensional dilute suspension of disks,
\begin{equation}
\eta^*=\eta\left(1+2\phi\right).
\label{eq:disks}
\end{equation}
Nevertheless, in experiments a reduction in viscosity is anticipated  \cite{sokolov1}.  In reality, bacteria are not spherical.  An interaction with the background flow thus produces a torque on the bacteria which will tend to align them in particular preferred directions (see e.g. \cite{pedley}).  For a fully  aligned suspension of bacteria, one obtains, from (\ref{evformula}),
\begin{equation}
\eta^*=\eta\left[1+2\phi+\frac{c}{2\pi\eta\epsilon_0}\frac{1+2\lambda(2+\lambda)}{(1+\lambda)^3}\phi\sin 2\alpha\right].
\end{equation}
For the background flow (\ref{bgflow}), the corresponding preferred directions are $\alpha=-\frac{\pi}{4},\frac{3\pi}{4}$.  Thus, after aligning,
\begin{equation}
\eta^*=\eta\left[1+2\phi-\frac{c}{2\pi\eta\epsilon_0}\frac{1+2\lambda(2+\lambda)}{(1+\lambda)^3}\phi\right].
\end{equation}
This formula can be used to explain the experimental data on the reduction of viscosity in bacterial suspensions observed in \cite{sokolov+}.

\section{Conclusion}\label{sec-conclusion}

Here a rather puzzling result has been obtained: the  effective viscosity of active bacterial suspension
{\it may decrease} with the increase in volume fraction of particles. Moreover, the viscosity may even become formally negative (when the flagella add more
energy than is being dissipated)
if the volume fraction $\phi$ and the magnitude of the propulsion force $c$ exceed critical values.  Manifestations of negative viscosity have been observed in
experiments \cite{sokolov+} via large-scale instability and the formation of non-decaying whorls and jets of collective locomotion (see also simulations \cite{graham,Ishikawa} for dense suspensions). The reduction of effective viscosity can be interpreted as a result of transformation by swimming bacteria of chemical energy of the surrounding nutrient medium into mechanical energy of fluid motion, and thus replacing energy loss due to viscous dissipation.

It is also notable that the coefficient $\epsilon_0$, which characterizes the strength of the background flow, shows up in the formula for effective viscosity (unlike in a Newtonian fluid).  Thus, the additional term due to self-propulsion is inherently non-Newtonian.  Since $\epsilon_0$ appears in the denominator, there is a blow up in the effective viscosity as $\epsilon_0\rightarrow 0$.  Nevertheless, this is to be expected, as the viscosity is, roughly, the ratio of stress of the entire flow to rate of strain of the background flow--in the case of a bacterial suspension, as the rate of strain of the background flow goes to zero, the stress does not, since the bacteria will continue swimming.

We also comment that the limit of vanishing background flow  $\epsilon_0 \to 0$  is not trivial. As mentioned above, the reduction of viscosity occurs only
when the bacteria become aligned in a certain direction determined by the principal axis of the strain-rate tensor. However, when $\epsilon_0\rightarrow 0$, the bacteria no longer have a tendency to align (also, a preexisting aligned state would be destroyed by fluctuations, e.g. due to random tumbling of bacteria) and hence the third term in Eq. (\ref{evformula}) does not blow up. Thus, in the limit of $\epsilon_0 \to 0$,  the steady state is in fact the isotropically oriented suspensions $X(\alpha)= 1/2\pi$, which recovers the classical result for disks Eq. (\ref{eq:disks}). Moreover, since  the effective viscosity for small strain rates is determined by the interplay between the response of the bacterium, through its orientation, to the shear strain and random fluctuations,  one needs to solve self-consistently the equation for the orientation distribution function $X(\alpha$) in the presence of non-zero strain rate.  This isotropic orientation can also explain the results of recent simulations
\cite{Ishikawa1} in which no reduction of the effective viscosity due to swimming bacteria (modeled by spherically symmetric particles) was detected: the reduction of viscosity requires particle reorientation by shear flow. However, the orientation of spherical particles is not affected by shear flow. In contrast, the effect is expected for elongated particles (ellipsoids or cylinders with large aspect ratios (of the order of 1:5  for swimming bacteria).

Obviously, more dedicated controlled experiments with bacterial suspensions and further generalizations of the obtained result to more general flow geometries, such as three-dimensional films and slabs, are keenly needed.
The puzzling phenomenon of reduced viscosity in bacteria-laden fluids may find rather unexpected  technological applications in bio-medical research and chemical technology, such as microscopic bacterial mixers and chemical reactors \cite{kim1}.

{\bf Acknowledgments}

The work of L. Berlyand was partially supported by NSF grant DMS-0708324. I.S. Aranson and D. Karpeev were supported by US DOE,
grant DOE grant DE-AC02-06CH11357.  

\appendix

\section{Calculation of Dissipation Rate and Definition of Effective Viscosity}\label{app-def_ev}
Following \cite{batchelor}, the rate of work being done at the boundary $\Gamma=\partial\Omega$ is given by
\begin{equation}
\int_{\Gamma}\sigma_{ij}u_i n_j dA=e_{ik}\int_{\Gamma}(-P\delta_{ij}+2\eta e_{ij}) x_k n_j dA,
\end{equation}
where $\sigma_{ij}$ is the stress tensor, $u$ is the velocity of the fluid, $n$ is the unit outward normal for the surface $\Gamma$, $P$ is the pressure of the fluid, and $e_{ij}$ is the rate of strain tensor.  Henceforth, primed quantities will denote the disturbance values due to the presence of a bacterial suspension.  The new stress tensor can be expressed as $\tilde{\sigma}_{ij}=\sigma_{ij}+\sigma'_{ij}$.  Additionally, on $\Gamma$, $e_{ij}=\epsilon_{ij}$.  Thus, the effective viscosity of the suspension $\eta^{*}$ is defined by setting
\begin{eqnarray}
\lefteqn{\epsilon_{ik}\int_{\Gamma}(-P\delta_{ij}+2\eta^{*} \epsilon_{ij}) x_k n_j dA=} \\
& \epsilon_{ik}\int_{\Gamma}(-P\delta_{ij}+2\eta \epsilon_{ij}+\sigma'_{ij}) x_k n_j dA. \nonumber
\end{eqnarray}
Noting that the terms involving $P$ are identical on both sides and employing the divergence theorem yields
\begin{equation}\label{defev}
2A(\eta^*-\eta)\epsilon_{ij}\epsilon_{ij}=\epsilon_{ik}\int_{\Gamma}\sigma'_{ij}x_k n_jdA,
\end{equation}
where $A$ is the area of the surface $\Gamma$.  The right hand side of (\ref{defev}) is the additional rate of dissipation due to the suspension.  Employing the divergence theorem yet again, this integral is transformed into an integral over the surfaces of the particles, producing
\begin{eqnarray}
%\begin{split}
\lefteqn{\epsilon_{ik}\int_{\Gamma}\sigma'_{ij}x_k n_jdA=} \\
&\epsilon_{ik}\int_{\Omega-\sum V_B\cup V_\epsilon} \left(\frac{\partial\sigma'_{ij}}{\partial x_j}x_k+\sigma'_{ik}\right)dV \nonumber \\
&+\epsilon_{ik}\sum\int_{\partial V_B\cup \partial V_\epsilon} \sigma'_{ij} x_k n_j dA, \nonumber
%\end{split}
\end{eqnarray}
where $V_B$ is the volume occupied by a single bacterium, $V_\epsilon$ is a ball of radius $\epsilon$ around its corresponding point force, and the summation is taken over all bacteria inside $\Omega$.  Now, the fluid in $\Omega$ obeys the inhomogeneous Stokes equation
\begin{equation}\label{stokesfull}
\left\{
\begin{array}{rlcc}
\eta\Delta\vec{u}'&=\nabla p'+\sum_l\vec{c}_l\delta_{\vec{r}_l} &&\text{in }\Omega-\sum_l V_{B_l} \\
\nabla\cdot\vec{u}'&=0 &&\text{in }\Omega-\sum_l V_{B_l} \\
\vec{u}'&=\vec{v}_l &&\text{on }\partial V_{B_l}\\
\vec{u}'&=\epsilon_{ij}x_j &&\text{on }\partial\Omega
\end{array},
\right.
\end{equation}
where the subscript $l$ has been added to all quantities that can vary among the bacteria.  In particular, $\vec{v}_l$
is its velocity of the $l$th bacterium, $\vec{r}_l$ is the location of its point force, and $\vec{c}_l$ indicates the
strength and orientation of each point force.  Thus, $\frac{\partial\sigma'_{ij}}{\partial x_j}=0$ in $\Omega-\sum
V_B\cup V_\epsilon$.
Additionally,
\begin{eqnarray}
%\begin{split}
\lefteqn{\epsilon_{ik}\int_{\Omega-\sum V_B\cup V_\epsilon} \sigma'_{ik}dV} \\
&=\epsilon_{ik}\int_{\Omega-\sum V_B\cup V_\epsilon} 2\eta \frac{\partial u'_i}{\partial x_k}dV \nonumber \\
&=-\epsilon_{ik}\sum \int_{\partial V_{B}\cup \partial V_\epsilon} 2\eta u'_i n_k dA, \nonumber
%\end{split}
\end{eqnarray}
since the surface integral over $\Gamma$ vanishes.  Applying these observations to (\ref{defev}) produces
\begin{eqnarray}
\lefteqn{2(\eta^*-\eta)\epsilon_{ij}\epsilon_{ij}} \\
&=\frac{\epsilon_{ik}}{A}\sum_l\int_{\partial V_{B_l}\cup \partial V_{\epsilon,l}}\left(\sigma'_{ij}x_k n_j-2\eta u'_i n_k\right)dA. \nonumber
\end{eqnarray}.

\section{Velocity}\label{app-velocity}
Recall that the velocity field $\vec{u}$ solves the Stokes equation,
\begin{equation}\label{stokesappendix}
\left\{
\begin{array}{rl}
\eta\Delta\vec{u}&=\nabla p+\vec{c}\delta_{\vec{r}_0} \text{ in } \Omega \\
\nabla\cdot\vec{u}&=0 \text{ in } \Omega \\
\vec{u}&=0 \text{ on } \Gamma
\end{array}.
\right.
\end{equation}

For simplicity, it is assumed that $\vec{c}=c\hat{x}$ and $\vec{r}_0=(\lambda+1)\hat{x}$.  Since $\nabla\cdot\vec{u}=0$, the velocity can be expressed as the (2D) curl of a scalar stream function $\Phi(x,y)$.  This curl, which operates on scalar functions and produces a vector function, is defined by
$$
Curl_V\Phi(x,y)=\left(\begin{array}{c}\frac{\partial\Phi}{\partial y}\\ -\frac{\partial\Phi}{\partial x} \end{array} \right).
$$
The scalar curl $Curl_S$, which operates on vectors, is defined as
$$
Curl_S\vec{u}=\frac{\partial u_x}{\partial y}-\frac{\partial u_y}{\partial x}
$$
so that $Curl_S Curl_V \Phi(x,y)=\Delta\Phi(x,y)$.  Substituting $\vec{u}=Curl_V\Phi(x,y)$ into (\ref{stokes}) and taking the scalar curl of both sides yields the inhomogeneous biharmonic equation
\begin{equation}\label{biharmonicappendix}
\Delta^2\Phi(x,y)=Curl_S \frac{c}{\eta}\hat{x}\delta_{\vec{r}_0}=\frac{\partial}{\partial y} \frac{c}{\eta}\delta_{\vec{r}_0}
\end{equation}
with boundary conditions
\begin{equation}
\Phi=\frac{\partial\Phi}{\partial n}=0 \text{ on } \Gamma.
\end{equation}

The Green's function for the domain $\Omega$ with these boundary conditions is derived in the exact same fashion as that for the unit disc, and has a simple form as a function of the complex variables $z=x+iy$ and $\zeta=\xi+i\gamma$, taken from \cite{garabedian}:
\begin{eqnarray}
\lefteqn{G(z,\zeta)=\frac{1}{8\pi}\left|z-\zeta\right|^2\log\left|\frac{z-\zeta}{1-\bar{\zeta}z}\right|} \\
&+\frac{1}{16\pi}\left(\left|z\right|^2-1\right)\left(\left|\zeta\right|^2-1\right). \nonumber
\end{eqnarray}
This yields the solution formula
\begin{eqnarray*}
%\begin{split}
\Phi(x,y)=&\int_{\Omega}G(x+iy,\xi+i\gamma)\frac{\partial}{\partial \gamma} \frac{c}{\eta}\delta_{\vec{r}_0}(\xi,\gamma)d\xi d\gamma \\
=&-\frac{c}{\eta}\int_{\Omega}\frac{\partial}{\partial \gamma}G(x+iy,\xi+i\gamma)\delta_{\vec{r}_0}d\xi d\gamma
%\end{split}
\end{eqnarray*}
and hence
\begin{equation}
\Phi(x,y)=-\frac{c}{\eta}\frac{\partial G}{\partial \gamma}(x+iy,1+\lambda).
\end{equation}

\section{Green's Function}\label{app-green}
This derivation follows that in \cite{garabedian}, with the only difference being that, in this case, the domain is the outside of the unit disk $D$.  Nevertheless, the mathematical details are identical.  The fundamental solution of the biharmonic equation is
\begin{equation}\label{fundamentalsolution}
\Lambda(z,\zeta)=\frac{1}{8\pi}\left|z-\zeta\right|^2 \log\left|z-\zeta\right|,
\end{equation}
and the general solution of the biharmonic equation is
\begin{equation}\label{generalsolution}
\Upsilon(z)=2{\rm Re} \left\{\bar{z}\Phi(z)+\Psi(z)\right\},
\end{equation}
where $\Phi$ and $\Psi$ are arbitrary analytic functions.  Thus, the problem is to find a function of the form (\ref{generalsolution}) that cancels (\ref{fundamentalsolution}) and its normal derivative on $\partial D$.  To facilitate this, (\ref{generalsolution}) can be equivalently written as
\begin{equation}
\Upsilon(z)=2{\rm Re} \left\{\left[z^2-1\right]\Phi(z)+\Psi(z)\right\}.
\end{equation}
Let $\Gamma(z,\zeta)=\Lambda+\Upsilon$.  Then, the condition $\Gamma=0$ on $\partial D$ is equivalent to
\begin{eqnarray}
%\begin{split}
\lefteqn{{\rm Re} \left\{\Psi(z)\right\}=-\frac{1}{8\pi}(z-\zeta)(\bar{z}-\bar{\zeta})\log\left|z-\zeta\right|} \\
&={\rm Re} \left\{-\frac{1}{8\pi}(z-\zeta)\left(\frac{1}{z}-\bar{\zeta}\right)\log(1-\bar{\zeta}z)\right\}, \nonumber
%\end{split}
\end{eqnarray}
since $\bar{z}=\frac{1}{z}$ there.  Since the equation inside the braces is analytic in the unit disk, it must be $\Psi(z)$.  Additionally, the condition that $\Gamma=\frac{\partial\Gamma}{\partial n}=0$ on $\partial D$ implies $\frac{\partial\Gamma}{\partial x}=\frac{\partial\Gamma}{\partial y}=0$ there, and so
\begin{eqnarray}
%\begin{split}
\frac{\partial\Gamma}{\partial z}=&\bar{z}{\rm Re} \{\Phi(z)\}+\frac{1}{2}\Psi'(z)+\frac{1}{16\pi}(\bar{z}-\bar{\zeta}) \\
&+\frac{1}{16\pi}(\bar{z}-\bar{\zeta})\log(z-\zeta)(\bar{z}-\bar{\zeta})=0. \nonumber
%\end{split}
\end{eqnarray}
Substituting $\Psi$ yields
\begin{eqnarray}
\lefteqn{{\rm Re} \{\Phi(z)\}=} \\
&{\rm Re}\left\{\frac{|\zeta|^2-1}{16\pi}-\frac{1}{8\pi}\left(1-\frac{\zeta}{z}\right)\log(1-\bar{\zeta}z)\right\}, \nonumber
\end{eqnarray}
and hence, using $\bar{z}=\frac{1}{z}$ on $\partial D$ once more,
\begin{equation}
\Phi(z)=\frac{|\zeta|^2-1}{16\pi}-\frac{1}{8\pi}\left(1-\frac{\zeta}{z}\right)\log(1-\bar{\zeta}z).
\end{equation}
Thus,
\begin{eqnarray}
\lefteqn{\Gamma(z,\zeta)=\frac{1}{8\pi}\left|z-\zeta\right|^2 \log\left|\frac{z-\zeta}{1-\bar{\zeta}z}\right|} \\
&+\frac{1}{16\pi}(|z|^2-1)(|\zeta|^2-1).\nonumber
\end{eqnarray}

\section{Pressure}\label{app-pressure}
By taking the divergence of (\ref{stokes}), one obtains
\begin{equation}
\left\{
\begin{array}{rl}
\Delta p&=-\nabla\cdot c\hat{x}\delta_{\vec{r}_0}=-c\frac{\partial}{\partial x}\delta_{\vec{r}_0} \\
\nabla p\cdot\hat{n}&=\eta\Delta\vec{u}\cdot\hat{n} \text{ on } \Gamma
\end{array}.
\right.
\end{equation}

Thus $p(x,y)$ is the real part of some function $\Pi(z)$, which is holomorphic in $\Omega^c$, such that $\Pi'(z)=\eta(\Delta u_x-i\Delta u_y)$ on $\Gamma$.  In fact, viewing $u_x$ and $u_y$ as functions of $x$ and $y$,
\begin{eqnarray}
\label{pressureintegral}
\lefteqn{\Pi(z)=\eta\int_\gamma\left[\Delta u_x(X(\zeta),Y(\zeta))\right.} \\
&\left.-i\Delta u_y(X(\zeta),Y(\zeta))\right] d\zeta, \nonumber
\end{eqnarray}
where $X(z)=\frac{1}{2}\left(z+\frac{1}{z}\right)$, $Y(z)=\frac{1}{2i}\left(z-\frac{1}{z}\right)$, and $\gamma$ is any path from the origin to $z$.  It is clear that
$p={\rm Re} (\Pi)$ satisfies the compatibility condition and is harmonic, except where the integrand has singularities, so it remains to check that $\Delta{\rm Re} \Pi(z)=-c\frac{\partial}{\partial x}\delta_{\vec{r}_0}$.  Performing the integration in (\ref{pressureintegral}) gives
\begin{equation}
\label{analyticpressureappendix}
\Pi(z)=\frac{-c}{2\pi}\frac{a^2\lambda^2(2+\lambda)^2z}{(1+\lambda)(a(1+\lambda)-z)((1+\lambda)z-a)^2}.
\end{equation}
Taking the real part of (\ref{analyticpressureappendix}) and expanding it about $x=a(1+\lambda)$ yields
\begin{equation}
p(r,\theta)=c\frac{\cos(\theta)}{2\pi r}+O(1),
\end{equation}
thus, indeed, $\Delta p(r,\theta)=-c\frac{\partial}{\partial x}\delta_{\vec{r}_0}$.

\section{Disturbance flow}\label{app-disturbance}
In terms of the stream function of the background flow, $\Psi_0(r,\theta)=r^2(\cos^2\theta-\sin^2\theta)$, the stream function for the disturbance flow $\vec{u}_d$, from \cite{circlethm}, is given by
\begin{eqnarray}
%\begin{split}
\lefteqn{\Psi=\frac{r^4-2r^2a^2}{a^4}\Psi_0\left(\frac{a^2}{r},\theta\right)} \\
&+\frac{r^3}{a^4}(r^2-a^2)\frac{\partial}{\partial r}\Psi_0\left(\frac{a^2}{r},\theta\right) \nonumber \\ &-\frac{(r^2-a^2)^2}{4a^4}\nabla^2\left[r^2\Psi_0\left(\frac{a^2}{r},\theta\right)\right]. \nonumber
%\end{split}
\end{eqnarray}
This yields the velocity
\begin{equation}
\vec{u}_d=\left(
\begin{array}{c}
-\frac{1}{r}\frac{\partial\Psi}{\partial\theta}\hat{r}\\
\frac{\partial\Psi}{\partial r}\hat{\theta}\end{array}\right).
\end{equation}
The corresponding pressure $p_d$ is, once more, the real part of
\begin{eqnarray}
\lefteqn{\Pi(z)=\eta\int_\gamma\left[\Delta u_{d,x}(X(\zeta),Y(\zeta))\right.} \\
& \left. -i\Delta u_{d,y}(X(\zeta),Y(\zeta))\right] d\zeta. \nonumber
\end{eqnarray}
Performing this integration gives
\begin{equation}
p_d(x,y)=-\frac{16a^2\eta x y}{r^4}.
\end{equation}

\end{document}